\documentclass[journal]{IEEEtran}
\usepackage{graphicx}

\begin{document}
\title{Design and Construction of Large Size Micromegas\\
Chambers for the Upgrade\\
of the ATLAS Muon Spectrometer}
%
% author names and IEEE memberships
% note positions of commas and nonbreaking spaces ( ~ ) LaTeX will not break
% a structure at a ~ so this keeps an author's name from being broken across
% two lines.
% use \thanks{} to gain access to the first footnote area
% a separate \thanks must be used for each paragraph as LaTeX2e's \thanks
% was not built to handle multiple paragraphs
%

\author{Philipp~L\"osel
        and~Ralph~M\"uller,~\IEEEmembership{on~behalf~of~the~ATLAS~Muon~Collaboration}% <-this % stops a space
\thanks{Manuscript received \today.} %This work was supported in part by the U.S. Department of Commerce under Grant No. BS123456 (sponsor acknowledgment goes here).}% <-this % stops a space
\thanks{Ph. L\"osel, Ludwig-Maximilians-Universit\"at, Munich, Germany (telephone: +49 89 289 14160, e-mail: philipp.loesel@physik.uni-muenchen.de).}%
\thanks{R. M\"uller, Ludwig-Maximilians-Universit\"at, Munich, Germany (telephone: +49 89 289 14160, e-mail: ralph.mueller@physik.uni-muenchen.de).}%
}

\maketitle
\pagestyle{empty}
\thispagestyle{empty}

\begin{abstract}
Large area Micromegas detectors will be employed for the first time in high-energy physics experiments. 
A total surface of about $\mathbf{150~m^2}$ of the forward regions of the Muon Spectrometer of the ATLAS detector at LHC will be equipped with 8-layer Micromegas modules. 
Each layer covers more than $\mathbf{2~m^2}$ for a total active area of $\mathbf{1200~m^2}$. 
Together with the small strip Thin Gap Chambers they will compose the two New Small Wheels, which will replace the innermost stations of the ATLAS endcap muon tracking system in the 2018/19 shutdown.

In order to achieve a 15$\mathbf{\%}$ transverse momentum resolution for $\mathbf{1~TeV}$ muons, in addition to an excellent intrinsic resolution,
the mechanical precision of each plane of the assembled module must be as good as $\mathbf{30~\mu m}$ along the precision coordinate and $\mathbf{80~\mu m}$ perpendicular to the chamber.

The design and construction procedure of the Micromegas modules will be presented, as well as the design for the assembly of modules onto the New Small Wheel.
Emphasis will be on the methods developed to achieve the challenging mechanical precision.

Measurements and simulations of deformations created on chamber prototypes 
as a function of thermal gradients, 
internal stress (mesh tension and module fixation on supports) 
and gas over-pressure  
%will be shown in comparison to simulation. These tests 
were essential in the development of the final design.

% in order to minimize the effects of deformations.\\
During installation and operation all deformations and relative misalignments will be monitored by an optical alignment system and 
compensated in the tracking software.
\end{abstract}

%\begin{IEEEkeywords}
%IEEEtran, journal, \LaTeX, paper, template.
%\end{IEEEkeywords}

\section{Introduction}
	\label{intro}

\begin{figure}[h!]
	\centering
	\includegraphics[width=3.in]{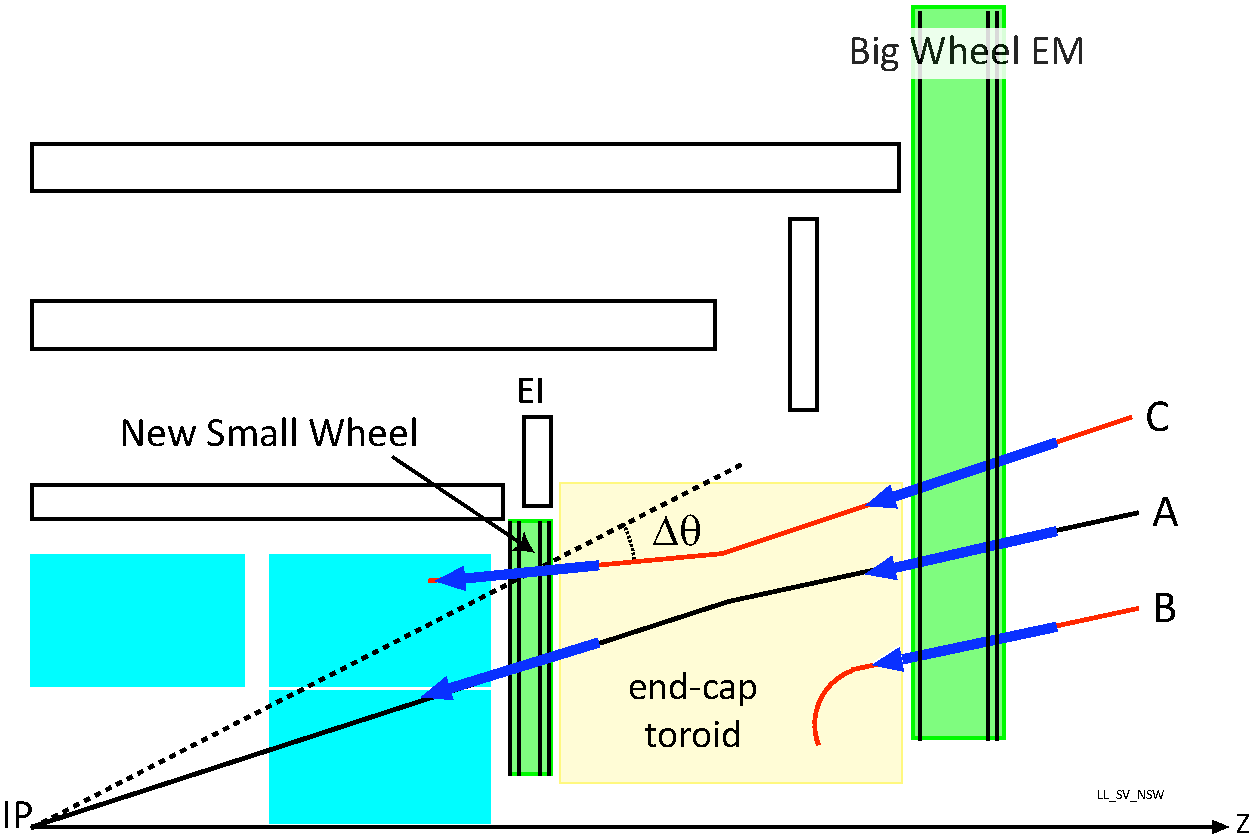}
	\caption{Schematic drawing of a quarter of the ATLAS detector 
	with possible tracks in the endcap region. 
	Track A leaves a track in all detector regions and points 
	towards the interaction point.
	Track B gives only a signal in the Big Wheel and Track C 
	does not point towards the interaction point. 
	Only Track A corresponds to an acceptable muon.}
	\label{atlasQuater_FIG}
\end{figure}

\IEEEPARstart{A}{fter} the long shut down 2 (2018/19) and the 
related upgrade of the Large Hadron Collider (LHC) at CERN  
the luminosity will reach values around 
$2\cdot\,10^{34}\,cm^{-2}s^{-1}$. 
The number of primary vertices 
per bunch crossing and the background rate will increase proportionally.  
As a consequence the background hit rate in the innermost endcap region of 
the muon spectrometer of the ATLAS detector \cite{atlas} (Small Wheel) 
will reach values above $10~kHz\,cm ^{-2}$, whereas
the currently installed precision tracking detectors   
will become ineffective at hit rates of $\approx2~kHz\,cm ^{-2}$. 
For this reason the detectors in the inner endcap region 
will be replaced by large size Micromegas 
(MICRO MEshed GAseous Structure \cite{MM_orig}) and 
sTGC (small Thin Gap Chamber \cite{TGC_orig}) detectors. 
While the sTGC detectors will be used mainly for triggering, 
the Micromegas detectors are foreseen mainly as precision tracking chambers.
For redundancy both detector types are designed to take over 
the other task as well. 
In this case the goal of the Micromegas would be to provide information to the trigger system \cite{atlas_trigger}, 
whether the traversing muon originated at the interaction point or not.
In the present setup of the ATLAS muon spectrometer 
only the middle part (Big Wheel) of the three endcap  
regions  is read out on trigger level including interaction point tracking. 
Fig. \ref{atlasQuater_FIG} shows three
possible track constellations which lead to an accepted muon track 
in this region. 

Combining the information of the Small Wheel and 
the Big Wheel a clear distinction between 
a muon coming from the interaction point and 
background hits will be possible. For this an angular resolution better than 1\,mrad is necessary in the NSW.

\begin{figure}[h!]
	\centering
	\includegraphics[width=3.in]{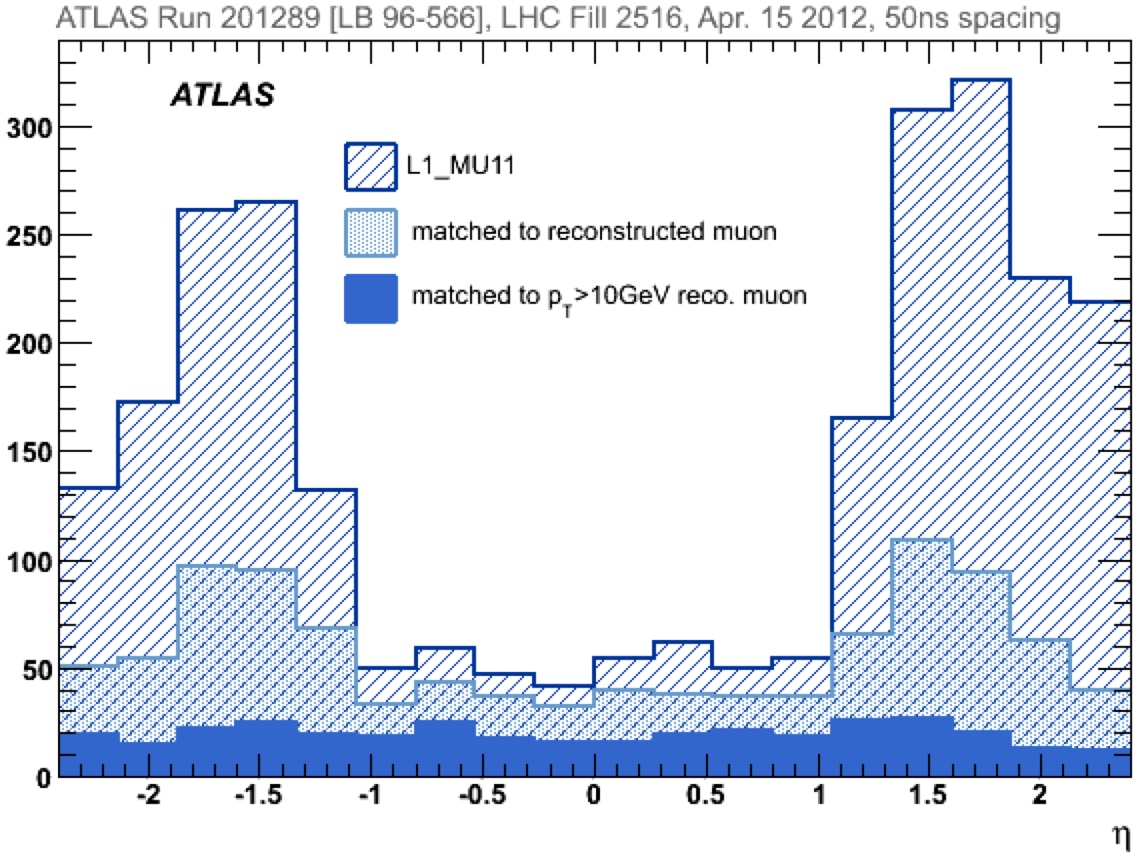}
	\caption{Trigger rate in the ATLAS detector 
	(April 2012) \cite{NSW_TDR}. The rate in the endcap region 
	$\left(\left|\eta\right|>1\right)$ is dominated by fake triggers. 
	Counting only real muon events with p$_t\,>\,1$\,TeV, 
	the trigger rate is flat in 
	the whole detector (solid blue histogram).}
	\label{muonTrigger_FIG}
\end{figure}

Extrapolating the current trigger rate in the endcap,
see fig.~\ref{muonTrigger_FIG},
to the luminosity after the second long shutdown 
indicates that the rate will be close to the bandwidth limit. 
%A plot of this scenario is shown in fig. \ref{muonTrigger_FIG}. 
The reason for this high trigger rate is caused by 
background events imitating the signals of true muons. 
Including the small wheel in the level one trigger, 
the total rate is expected to reduce by a factor of $5$.

\section{Working Principle of Resistive Strip Micromegas Detectors}
	\label{WPoMD}

\begin{figure}[h]
	\centering
	\includegraphics[width=3.5in]{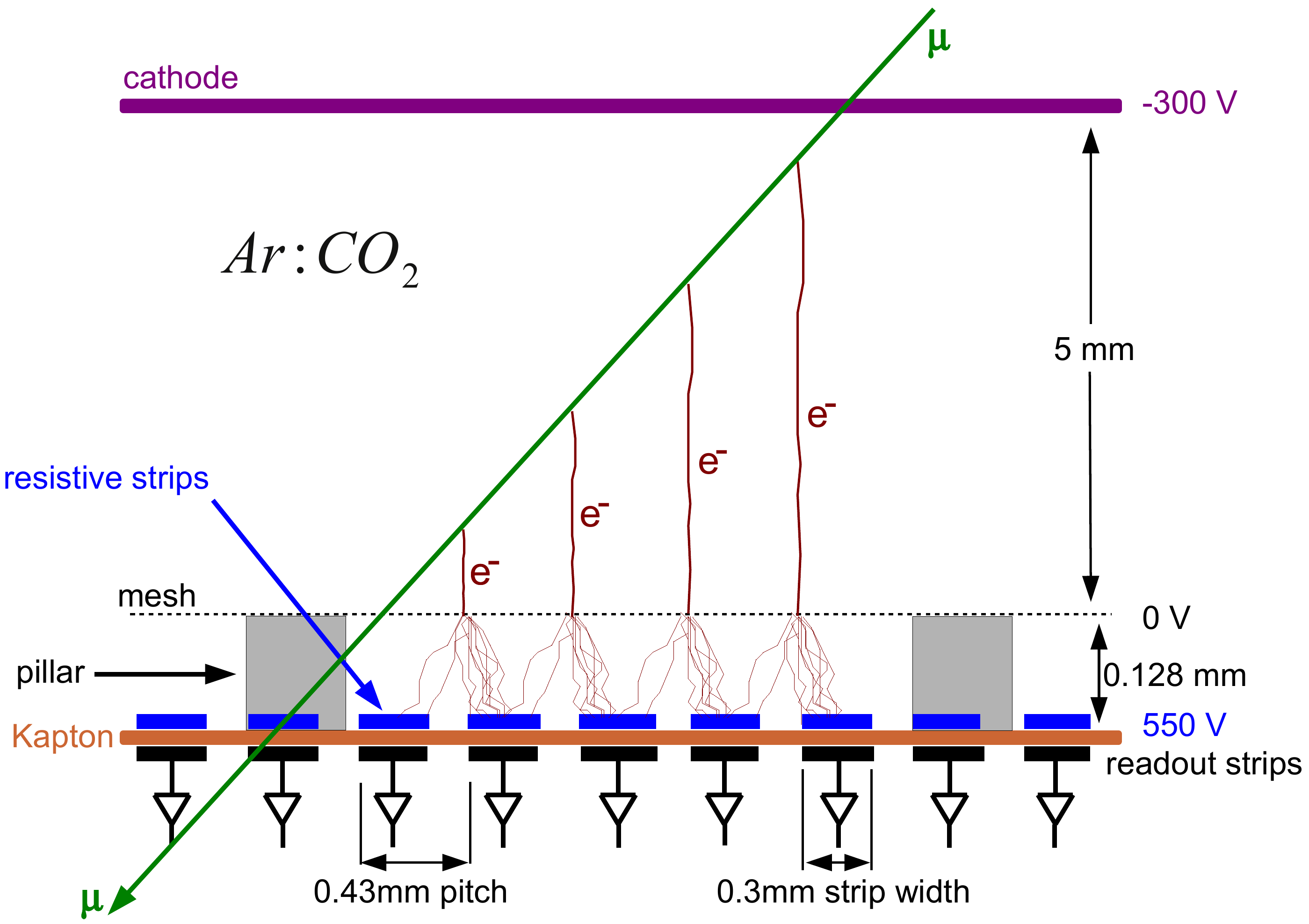}
	\caption{Sketch of charged particle detection using a Micromegas detector. 
	A muon (green) traverses the detector ionizing the Ar:CO$_2$ 93:7 gas. 
	The electrons (red) drift to the mesh and cause avalanches between 
	the mesh and the anode strips. The accumulated charge is 
	coupled capacitively to the readout strips underneath the anode strips.}
	\label{MM_workPrincip_FIG}
\end{figure}

Micromegas detectors are high rate capable precision tracking devices. 
They consist of three planar structures: a 
copper cathode, a stainless steel micro mesh and anode strips
on printed circuit boards (PCBs). 
As shown in fig. \ref{MM_workPrincip_FIG} a 
traversing charged particle ionizes the gas along its path through the detector.

In the drift region the electron-ion pairs are separated due to 
the electrical field between the cathode and the mesh 
%At electric fields 
of around $E_{drift}\approx 600~Vcm^{-1}$. 
The electrons drift towards the mesh and into the amplification region 
where an avalanche is forming by secondary ionisation in the 
narrow region between mesh and resistive strips of $d_{amp}=128~\mu m$. 
Typical amplification fields are here $E_{amp}\approx43~kVcm^{-1}$ leading to a gas gain of a few $1000$. 
Discharges between mesh and strips can occur when the 
charge density in the 
high field in this region exceeds 
the Raether limit of 2$\times 10^8 ~ e ~ mm^{-2}$ \cite{raether},
e.g. when a strongly ionizing ion enters a detector 
having been set up for detection of minimum ionizing muon. 
This is non-destructive 
but causes a temporarily reduced amplification field
and thus deadtime. 
At standard micromegas the 
whole detector gets insensitive until the field is re-established,
as the potential between the anode and the mesh equilibrates. 
To prevent this global voltage drop a layer of 
resistive strips is added on top of the readout strips 
with a resistivity of $\approx10~M\Omega~cm^{-1}$ \cite{theo}.
The local discharge affects only one or a few strips and 
the high resistivity of the anode strips 
allows for an equilibration of the local potentials 
on a tiny region only and thus the fast breakdown of the discharge 
streamer and a fast recharge process of the anode.
Only a negligible part of the detector is then affected by such a 
discharge. 
For electronic readout the
charge collected on the resistive strips is 
capacitively coupled to the readout strips. 
They have a width of $=300~\mu m$ and a pitch of 
$425~\mu m$ or $450~\mu m$ depending on the module
in the New Small Wheel. 
The homogenous tiny distance between anode and mesh is 
established by a regular structure of pillars made lithographically of 
Pyralux coverlay.

\section{New Small Wheel Sector Design}
	\label{NSWSD}

\begin{figure}[h]
	\centering
	\includegraphics[width=3.in]{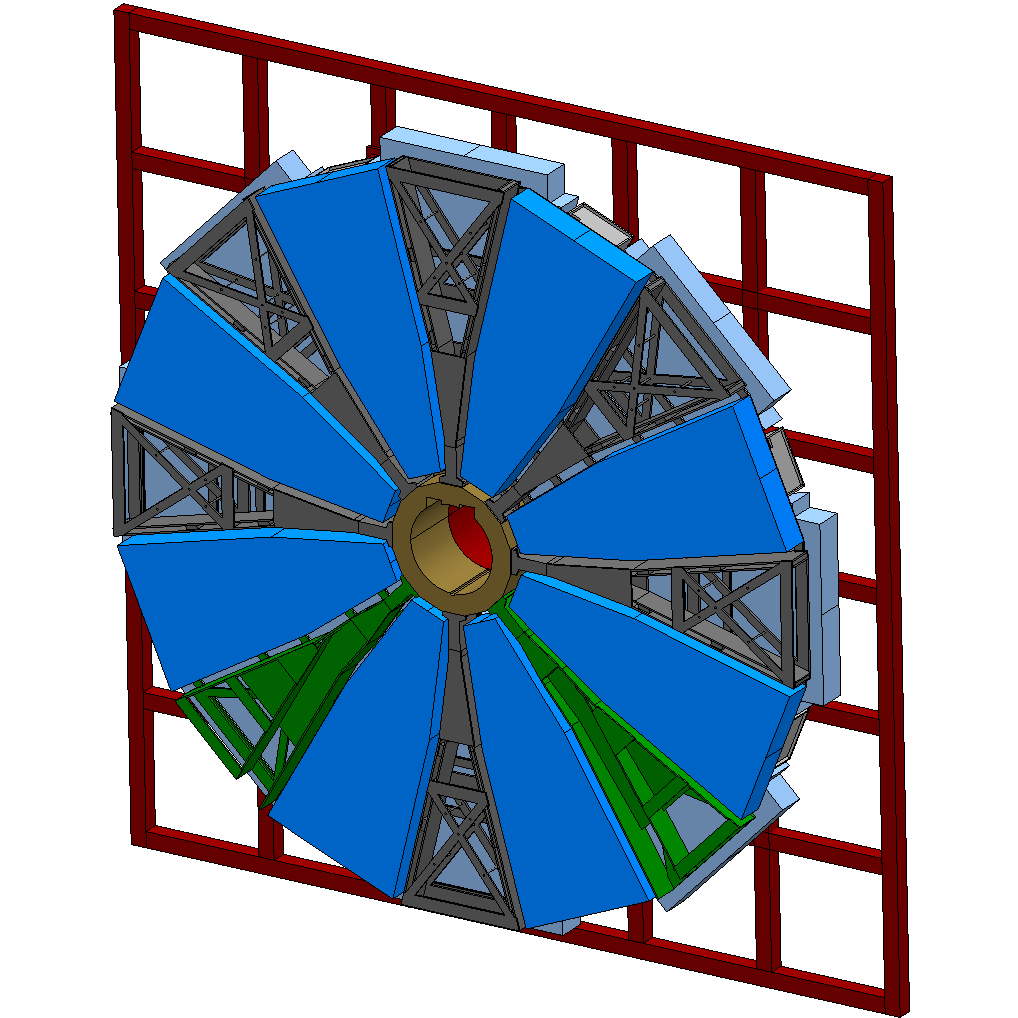}
	\caption{The New Small Wheel with eight 
	alternating small (dark blue) and large (light blue) sectors. 
	To cover the whole region of the NSW the sectors overlap partially.}
	\label{nswSector_FIG}
\end{figure}

The New Small Wheel (NSW) 
%consisting of Micromegas and sTGCs 
is designed to have the identical 
segmented structure like the current Small Wheel. 
It will be subdivided in eight alternating small and 
large sectors as shown in fig. \ref{nswSector_FIG}.

\begin{figure}[h]
	\centering
	\includegraphics[width=3.5in]{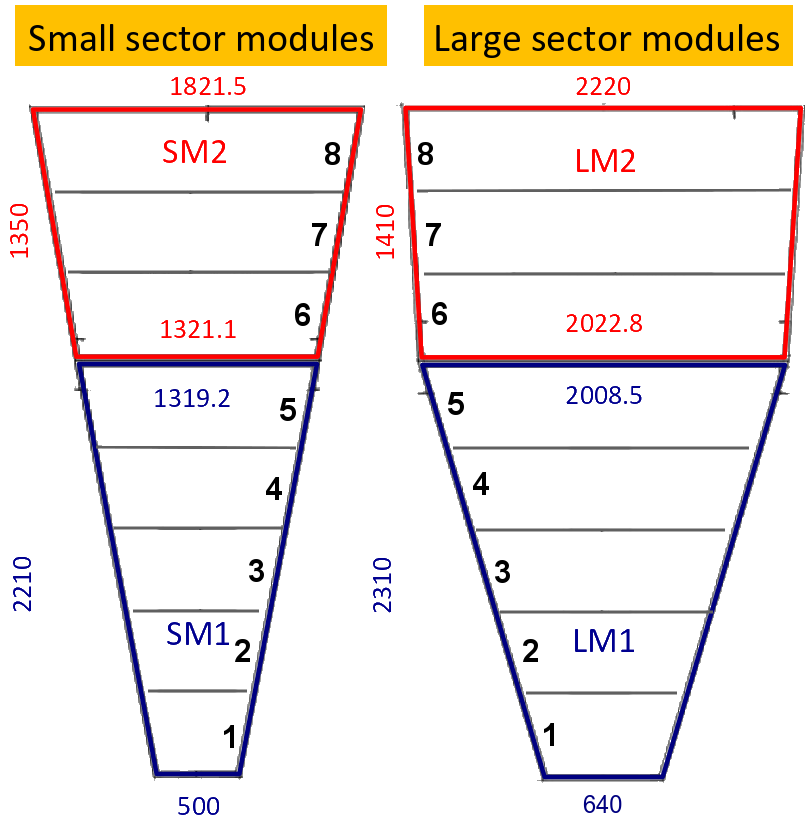}
	\caption{Small and large sectors of the NSW divided into 
	four trapezoidal modules with different dimensions. 
	All units are in mm.
	For each sector the active readout plane is subdivided into
	eight readout PCBs and two detector modules.}
	\label{sectordimensions_FIG}
\end{figure}

For the Micromegas each sector will be subdivided into two subsectors. 
There will be four trapezoidal types of Micromegas modules 
with different dimensions 
(see fig. \ref{sectordimensions_FIG}), which are built and 
assembled in four different construction sites,
see table \ref{constSites_TAB}.

\begin{table}[h]
	\centering
	\begin{tabular}{|c|l|l|}
	\hline
		SM1 & Italy \\ 
		\hline 
		SM2 & Germany \\ 
		\hline 
		LM1 & CERN, Greece and Russia \\ 
		\hline 
		LM2 & France \\ 
		\hline 
	\end{tabular} 
	\caption{Table of construction sites for the four Micromegas
	modules of the NSW project. 
	For details see TDR \cite{NSW_TDR}.}
	\label{constSites_TAB}
\end{table}

To reach an optimal track resolution each module consists 
of eight consecutive Micromegas detectors, which are formed 
by two quadruplets (see fig. \ref{quadruplet_FIG}). 
Each quadruplet is constructed of five sandwich panels, 
see section \ref{PCUsS} and \ref{VQAiaCR}. 
Two layers of a quadruplet have a strip configuration parallel 
to the upper and lower bound, so called $\eta$-strips. 
The other two layers have stereo strips tilted by $+1.5^{\circ}$ ($s$-strips) 
and $-1.5^{\circ}$ ($s'$-strips) with respect to the $\eta$ strips. 
The stereo strips deliver a rough information 
about the azimuthal $\varphi$-coordinate of the muon.
Between the double quadruplet structure a spacer frame will be installed 
for stable mounting. 
%This results in a final strip order for one sector: 
%$\eta-\eta-s-s'$ spacer $s-s'-\eta-\eta$.
% Das ist leider gar nicht klar.
 
A Micromegas sector is sandwiched by sTGCs.

\section{Panel Construction Using a Stiffback}
	\label{PCUsS}
In order to reach the desired position resolution of 
better than $100~\mu m$ in a single detector layer, 
the mechanical requirements of each panel are very strict.
The panels, sandwiches of FR4 PCB sheets 
with an aluminum
honeycomb core, need to have a surface planarity of 
better than $80~\mu m$. Even more strict is the requirement for 
readout-boards: the horizontal deviation of the readout strips position must not 
exceed $30~\mu m$ \cite{NSW_TDR}.

\begin{figure}[h]
	\centering
	\includegraphics[width=3.5in]{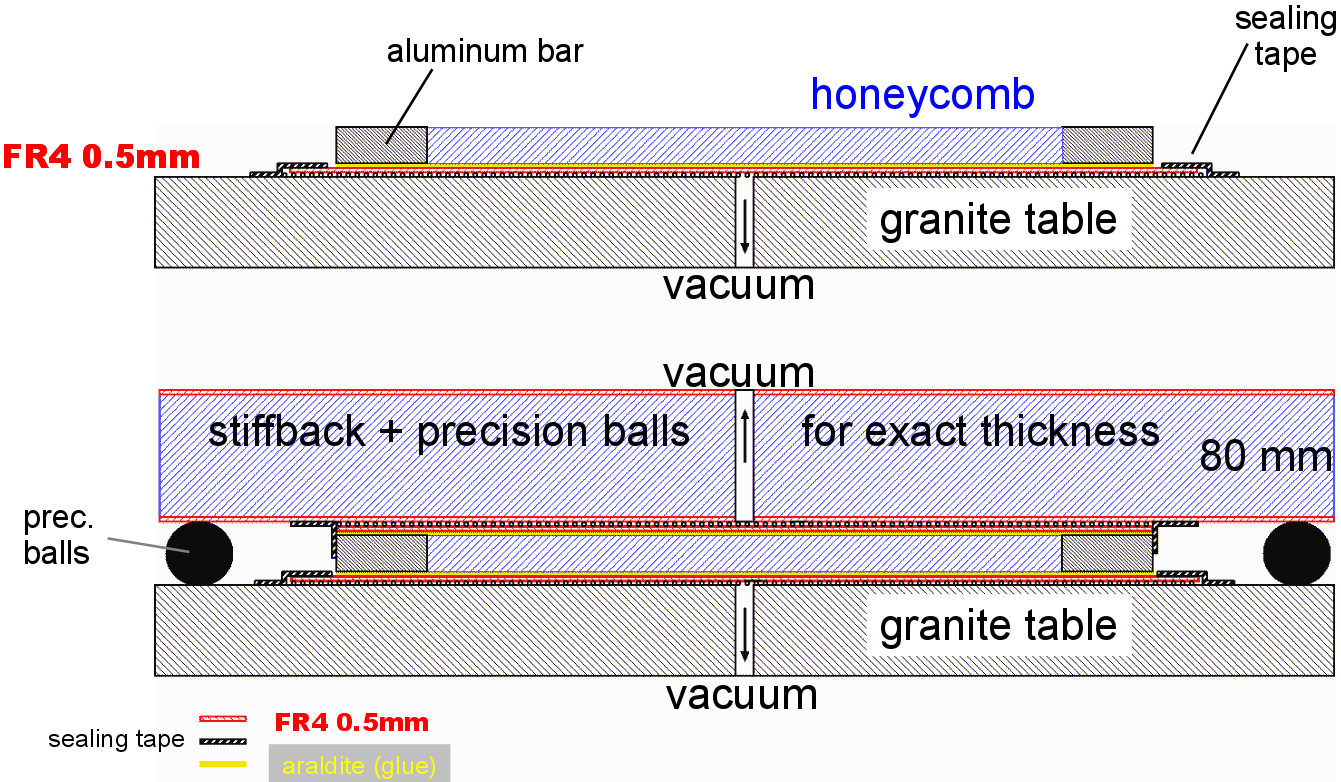}
	\caption{The two steps of glueing a sandwich by use of a stiffback. 
	The upper picture shows the final assembly of the first glueing step, 
	the lower part the second step.
	The lower planar surface of the stiffback contains many small
	holes for vacuum suction. The honeycomb is microperforated 
	and contains a few manually drilled channels.
	The distance between the stiffback and granite surface 
	is controlled by exact distance pieces like the balls from
	ball bearings shown here.}
	\label{twoStepGlueing_FIG}
\end{figure}

The four construction sites are using similar
methods for the panel constructing to reach 
the mechanical requirements.
Individual adaptions 
are accounted for by locally available 
infrastructure and manpower. 
The idea is to transfer the planarity 
of a very planar surface, e.g. a granite table, 
to PCBs by application of vacuum, see fig.~\ref{twoStepGlueing_FIG}.
A frame of extruded aluminum bars 
and a honeycomb structure    
are glued on top of the PCB
for stabilisation and reinforcement. 
A second PCB-layer, sucked down again to a planar surface,
is glued simultaneously or
in a second step to this PCB-aluminum~frame-honeycomb-structure.

Thus, two slightly different methods will 
be used to build the panels. In both ways at least one of 
the planar reference surfaces is a so called stiffback,
a stable and accurate holding structure.

\begin{figure}[h]
	\centering
	\includegraphics[width=3.5in]{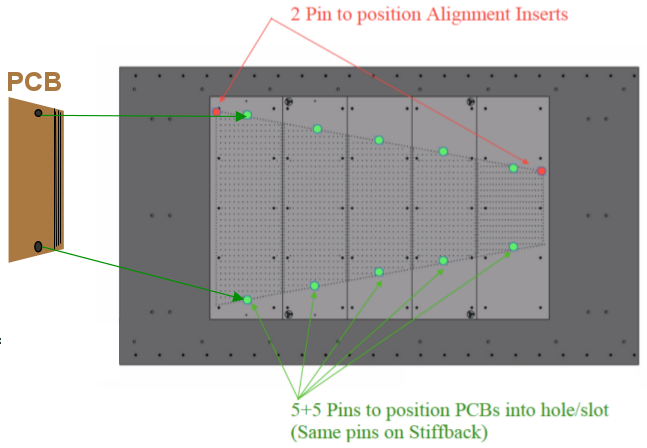}
	\caption{Readout PCBs are positioned on reference surfaces (light gray) made of precision surface massive vacuum suction ground plates.
	The stiffback, not shown here, is constructed similarly.
	Both reference surfaces enable exact positioning of readout PCBs via alignment pins and mutual alignment by commercial tapered interlocks.}
	\label{pavia_FIG}
\end{figure}

%and built as a negative of a granite table.
Stiffbacks will be either precision surface
massive aluminum vacuum suction ground
plates (see fig. \ref{pavia_FIG}) or relatively light sandwiches based on 
carbon fiber technology sheets or aluminum sheets,
both with aluminum honeycomb core,
built as a negative of a granite table.
The choice of the material is driven partly by 
available infrastructure.
A leightweight version of a stiffback consists of two 1mm thick
aluminum plates sandwiching an 80 mm thick honeycomb core.
For production an aluminum plate is sucked to a flat granite table and 
adopts its planar surface. 
Subsequently an $80~mm$ thick aluminum honeycomb is 
glued onto the plate to preserve the flatness of the plate. 
Finally a frame with vacuum suction ports 
surrounding the honeycomb and 
a second plate will be glued on top.
As glue the two component adhesive ARALDITE 2011 
is used for the stiffback as well as later for the panels.

The honeycomb is micro perforated 
with additional channels drilled 
manually before glueing
to allow for faster air exhaust.
To use the stiffback as a vacuum suction surface 
many small holes are drilled into the planar surface of the stiffback 
for vacuum buildup between the reference surface and the PCB sheet
to be attached.

The following explains both methods that will be used for panel
glueing, the single step glueing procedure 
or the two-step process.

	\subsection{Two Step Glueing Technique}
		\label{TSGT}
The method to build a panel is similar to the construction 
of the full aluminum stiffback. 
Here, each surface sheet of a panel will consist 
of three or five PCBs since single sheets in this size 
are not available (see fig. \ref{sectordimensions_FIG}). Aluminum honeycomb 
is used as sandwich core.
In the following we distinguish between drift- and readout-boards.

\begin{figure}[h]
	\centering
	\includegraphics[width=3.5in]{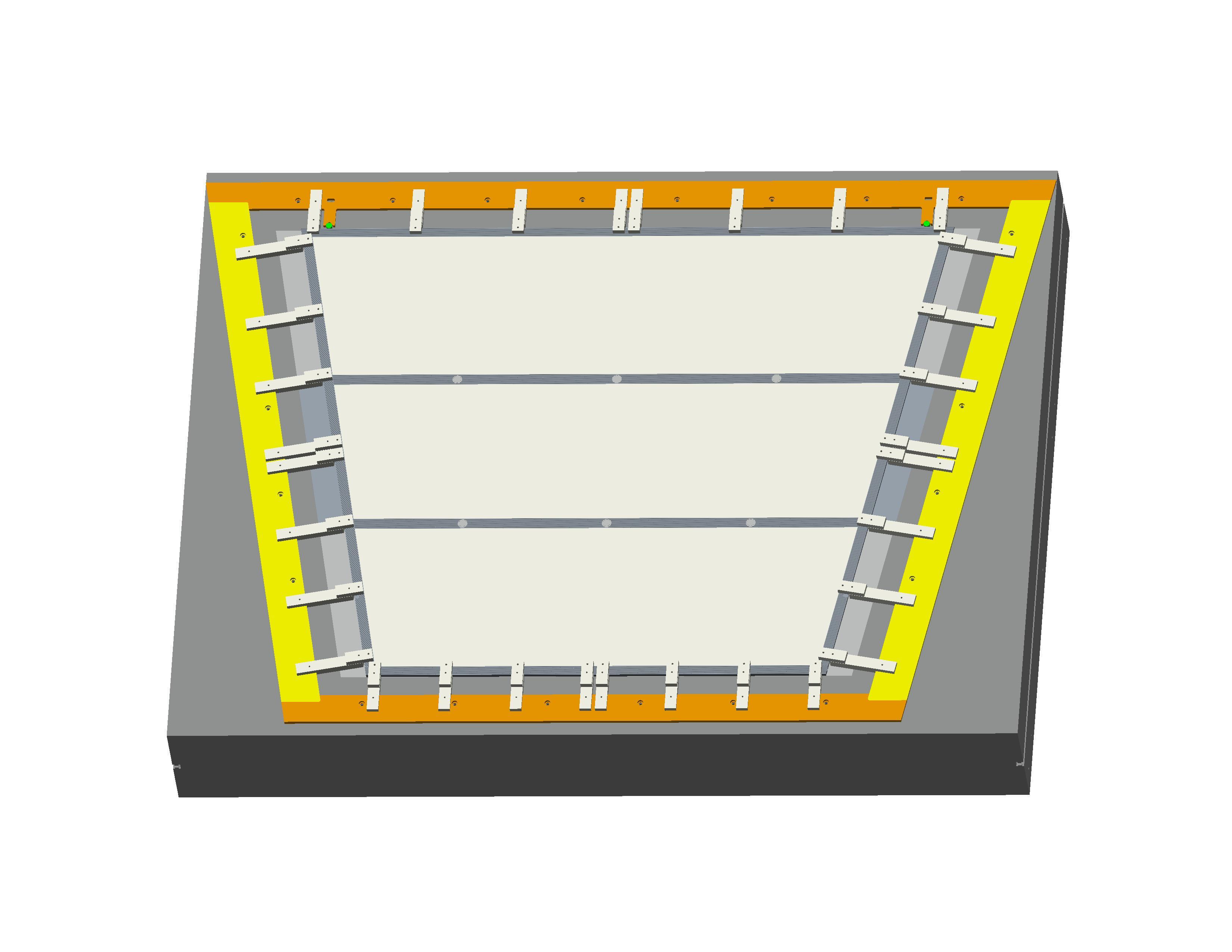}
	\caption{The external reference frame for module SM2 
	on the granite table. Exact distance fingers
	allow for the positioning of the drift PCBs,
	of the extruded aluminum frames and for the 
	alignment of the upper and lower PCB plane when the 
	stiffback is used.}
	\label{ext_ref_frame_FIG}
\end{figure}

The production of drift-panels is less sophisticated.
For the SM2 module all PCB-sheets forming a drift-plane will be positioned
on the planar reference table by use of  
exact distance pieces aligned against an external frame.
Sucking them down to the granite table fixes their position (see fig. \ref{ext_ref_frame_FIG}).
An aluminum frame 
of extruded aluminum bars is glued on top of the sheets. 
The aluminum frame has a height of $10~mm$. 
For reinforcement a $10~mm$ thick aluminum honeycomb is glued inside this frame. 
This first step of the glueing process is indicated in the upper part of 
fig.~\ref{twoStepGlueing_FIG}.

After curing, the product of the first glueing step 
is removed from the granite table
and sucked 
to a predefined position on the stiffback. 
The sheets for the second side of the panel are positioned
on the planar reference table. 
Finally the stiffback is turned upside down and the product of the 
first glueing step is glued to these PCB sheets.  
The relative position of both planes 
is hereby referenced again by exact distance pieces 
aligned to the external frame.

Precision distance pieces, balls in fig. \ref{twoStepGlueing_FIG},
between the granite table and the stiffback ensure 
the correct thickness and parallelism of the panel. 
The lower part of fig.~\ref{twoStepGlueing_FIG} shows the setup during 
curing of the second step.

	\subsection{Single Step Glueing Technique}
		\label{SSGT}
\begin{figure}[h!]
	\centering
	\includegraphics[width=3.5in]{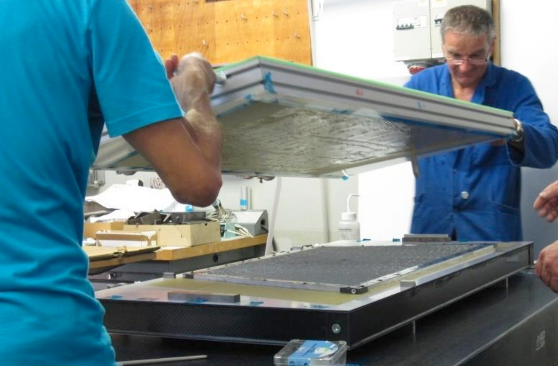}
	\caption{Glueing of a prototype panel in a single step process
	using the double stiffback method. 
	%Both stiffbacks contain channels and holes
	%for vacuum suction of PCB sheets to their surface.
	Two sets of PCB sheets  
	will be sucked to the respective stiffback surface 
	using small holes 
	in the surface sheets of the stiffback.
	The honeycomb and aluminum bars are then glued in between.}
	\label{MSW2_construction_FIG}
\end{figure}

%Slightly different to the first method some production sitesprefer a 
In the single step 
glueing technique two stiffbacks
with pins for mutual alignment and further pins for 
exact positioning of PCB sheets
are used instead of the granite table-stiffback combination. 
A photograph of the construction of a prototype can be seen in 
fig.~\ref{MSW2_construction_FIG}.

One set of PCB sheets each is sucked against the surface of 
the respective stiffback. The aluminum frame and the honeycomb
are then glued in a single step in between.
%The weight of the upper stiffback provides sufficient
%pressure during curing. 

	\subsection{Precise Alignment of the Readout Strips for the Glueing Process}
		\label{PAGP}
During production of readout sandwich-panels
the challenging requirement of the 
exact positioning of 
the readout-strips must be met and each sheet must 
be precisely aligned within 30 $\mu m$ to a global reference before 
sucking it to the granite table.

On each readout PCB two precise markers are produced 
in the same lithographic process as the readout strips.
These markers are positioned left and right of the readout strips. 
Their position is very accurately defined in regard to the 
position of the readout stips.
The markers are positioned vertically in the 
middle of the PCB.

For the single step glueing process 
precise alignment structures, 
a hole and an elongated hole to allow for thermal expansion,
will be drilled into the PCBs
at the position of the markers.  
The stiffbacks in fig.~\ref{MSW2_construction_FIG} contain 
matching alignment pins.

\begin{figure}[h!]
	\centering
	\includegraphics[width=3.5in]{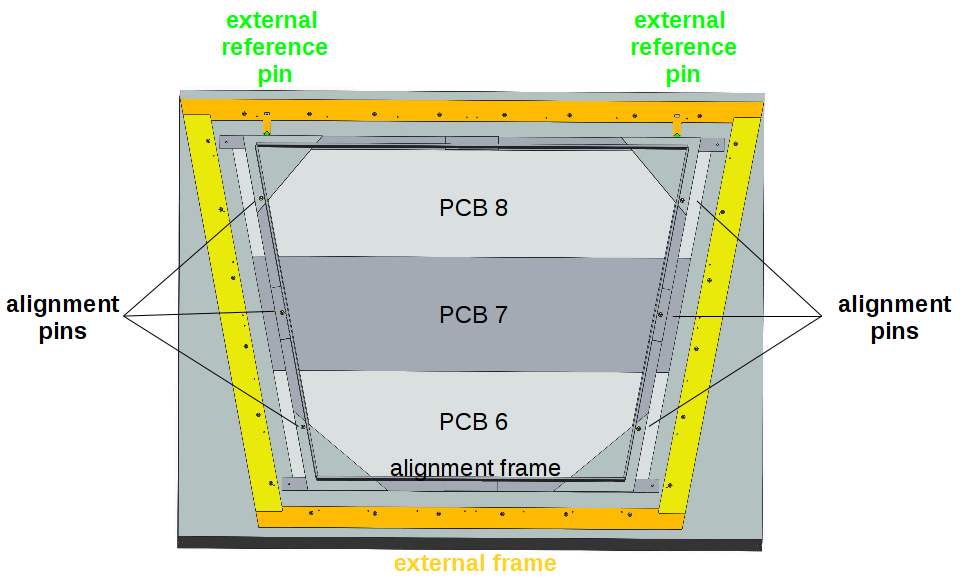}
	\caption{The alignment frame for the SM2 module.
	Two precision pins (green) define the position of the 
	readout panels on the granite table.
	They are mounted on an external permanently installed
	reference frame.
	The alignment frame attaches both pins by a V-shaped and a
	line-shaped contact piece respectively.
	Six 8\,mm precision pins provide guidance for the three 
	readout PCBs of the SM2 module. 
	The pins fit exactly into precision washers,
	a round one and an elongated one on opposing sides of the PCBs, 
	that are glued concentrically on top of 
	precision markers. These markers are produced in the same 
	lithographic step as the readout strips and 
	reference with high accuracy the position of the strips.
	A relative alignment 
	of the PCBs with an accuracy better than 30\,$\mu m$ is expected.
	}
	\label{MSW2_alignment_frame_FIG}
\end{figure}

For the two step glueing process
a precise washer containing 
a hole and a second one containing an elongated hole 
will be glued with very high precision and optically controlled 
on top of the markers.
For the SM2 module an alignment frame with six alignment pins
is used to position  
the PCBs by means of the washers  
on the granite table. As global reference 
serve two 16 mm diameter pins.
These are permanently installed on the
granite table. A V-shaped and line-shaped contact piece provide the 
correct position of the alignment frame against these external references.
For details see fig.~\ref{MSW2_alignment_frame_FIG}. 
The relative position of V-shaped and line-shaped contact pieces
against the six alignment pins is done in a single milling step
and will be controlled on a precise coordinate measurement machine.

The remaining glueing process is similar to the production
of the drift panels.

\section{Treatment of the Micromesh}
	\label{TotM}

%Another part of a Micromegas detector is the micro mesh. 
For the NSW detectors a stainless steel micromesh 
with a wire diameter of $28~\mu m$ and 
a distance between the wires of $50~\mu m$ is foreseen.
It will be mounted on the mesh frame of the drift panels. 
Precise $5.06~mm$ thick extruded
bars provide the correct distance between mesh and the cathode
surface.
The mesh will thus be attached to ground potential.
This so called floating mesh construction has the advantage that 
the amplification regions are still accessible even after the mounting of the 
meshes, an important feature for these large area detectors.
Before glueing the meshes all 
drift panels must be leak tested 
and all meshes and panels must be dust and dirt free.

\begin{figure}[h!]
	\centering
	\includegraphics[width=3.5in]{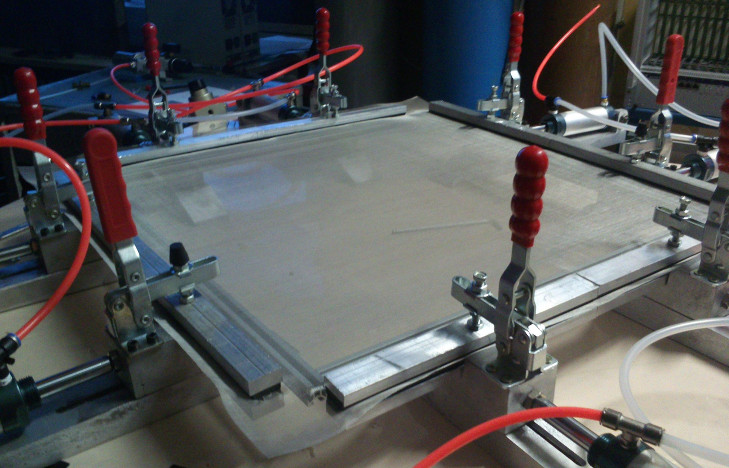}
	\caption{Prototype setup for stainless steel mesh streching. 
	A $60\times60\,cm^2$ mesh is stretched using 
	eight pneumatic clamps. For the final mesh size of the SM2 module 
	$24$ pneumatic 
	clamps will be needed.}
	\label{meshStreching_FIG}
\end{figure}

\begin{figure}[h!]
	\centering
	\includegraphics[width=3.5in]{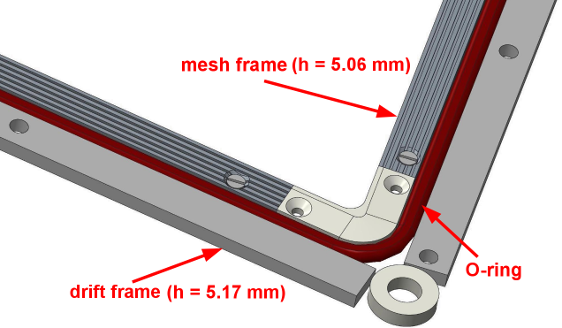}
	\caption{Detail of the mesh frame, which will be screwed and glued 
	on a drift panel. The mesh will be glued on top of the mesh frame. 
	An O-ring between mesh and drift frame will ensure gas tightness.
	}
	\label{meshFrame_FIG}
\end{figure}

The meshes will be streched with a homogeneous tension of $10~N\,cm^{-1}$ 
using pneumatic clamps, as shown in fig.~\ref{meshStreching_FIG},
and glued to transfer frames. This process can be centralized
and decoupled from the production sites for drift panels. 
The meshes will be transfered 
and glued in a later step to the mesh frame of a drift panel. 
The mesh frame 
consists of extruded aluminum bars 
with enlarged surface optimized for glueing, see fig.~\ref{meshFrame_FIG}.  

%The stretched mesh will be glued on a transfer frame. 
The extruded bars are screwed and glued to 
the cathode sides of the drift panels.
The additional glueing provides gas tightness between the active 
detector volume and the honeycomb volume.
The mesh is then glued to the mesh frame and the 
transfer frame can be reused after cleaning for the next mesh.\\
For the gas tightness an O-ring will be used in the groove
between the mesh frame and the outer drift frame.
The latter
defines the distance between anode and cathode and thus the drift region. 
The upper right corner of a SM2 frame 
is shown in fig. \ref{meshFrame_FIG}.

\section{Vertical Quadruplet Assembly in a Clean Room}
	\label{VQAiaCR}

\begin{figure}[h!]
	\centering
	\includegraphics[width=3.5in]{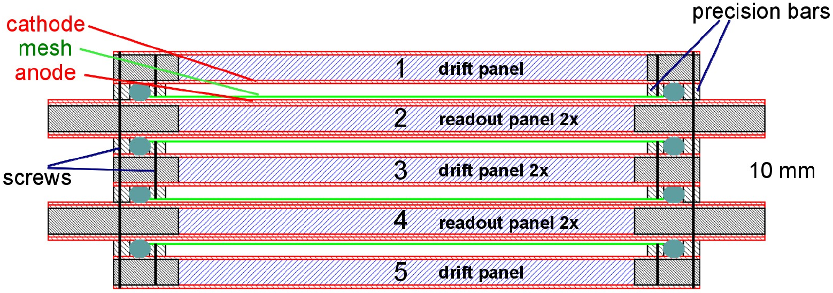}
	\caption{Scheme of the five sandwich panels 
	assembled to a quadruplet.
	The four active gas volumes between the
	drift and readout panels 
	are defined by extruded precision drift bars.
	Similarly, extruded mesh bars define the 
	exact distance of the micromeshes from the cathodes.
	In this design the meshes can be removed
	from the anode structures. Electrically they will be coupled to ground potential.}
	\label{quadruplet_FIG}
\end{figure}

%As already mentioned in section \ref{NSWSD} the NSW consists of Micromegas
%assembled of four active detectorlayers called quadruplets. 
%In fig. \ref{quadruplet_FIG} five of the previously discribed %precise sandwich 
%panels form a quadruplet.
%Three kinds of panels can be distinguished by their surfaces. 
%There are two panels with one copper drift cathode each. 
%They are the outermost panels.
%Two double readout panels act as anodes with congruent resistive and readout strips on both sides. 
%The panel in the center of the quadruplet has drift cathodes on both sides. \\
Fig.~\ref{quadruplet_FIG} shows a schematic cross section of a quadruplet.

\begin{figure}[h!]
	\centering
	\includegraphics[width=3.5in]{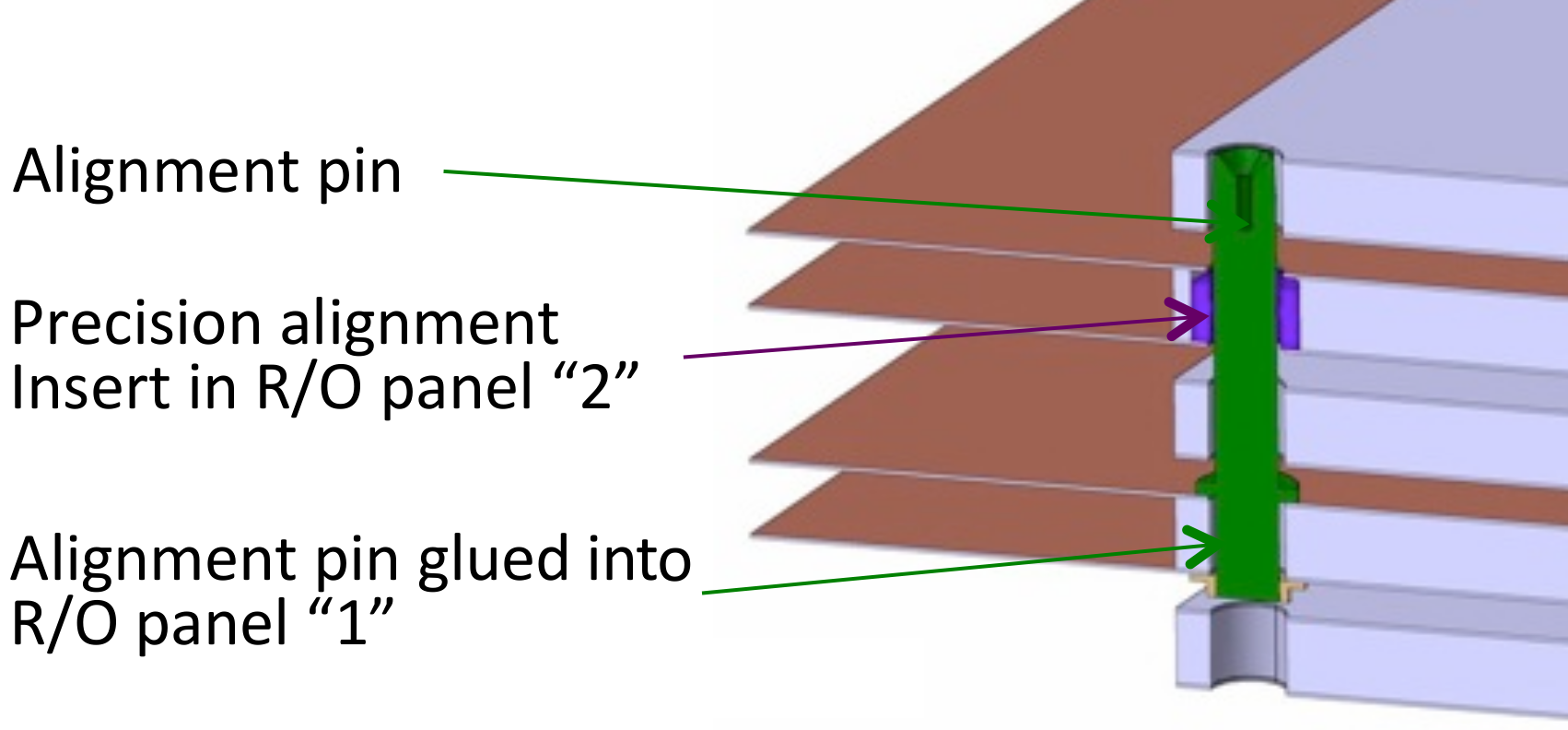}
	\caption{Precision pins are used for the alignment of the readout panels. 
	Two precision pins are glued to the lower readout panel.
	The upper readout panel has two precision inserts, 
	a round one and an elongated one, 
	to ensure the alignment
	between the readout panels.}
	\label{alignmentPin_FIG}
\end{figure}

%The meshes as described in section \ref{TotM} are mounted on the drift panels.\\
During the assembly of the five panels it is important 
that all layers are mounted parallel and that all strips 
in all four readout layers follow with highest accuracy 
their theoretical coordinates. 
This means that the two readout panels need to be precisely aligned. 
Figure \ref{alignmentPin_FIG} shows the method to achieve this using two  
precise alignment pins mounted to one of the readout panels and 
a round and an elongated precision insert in the second readout panel. 
% seite mit g6 und H7 erklaerung http://www.halder.de/halder_pdf/katalogseiten/1_DE/N4_DE_S421.pdf
The precision pins and inserts must have a relative clearance of 
$25~\mu m$ at the maximum.%(g6 and H7) (TODO was genau muss 25 micros sein?)

\begin{figure}[h!]
	\centering
	\includegraphics[width=3.in]{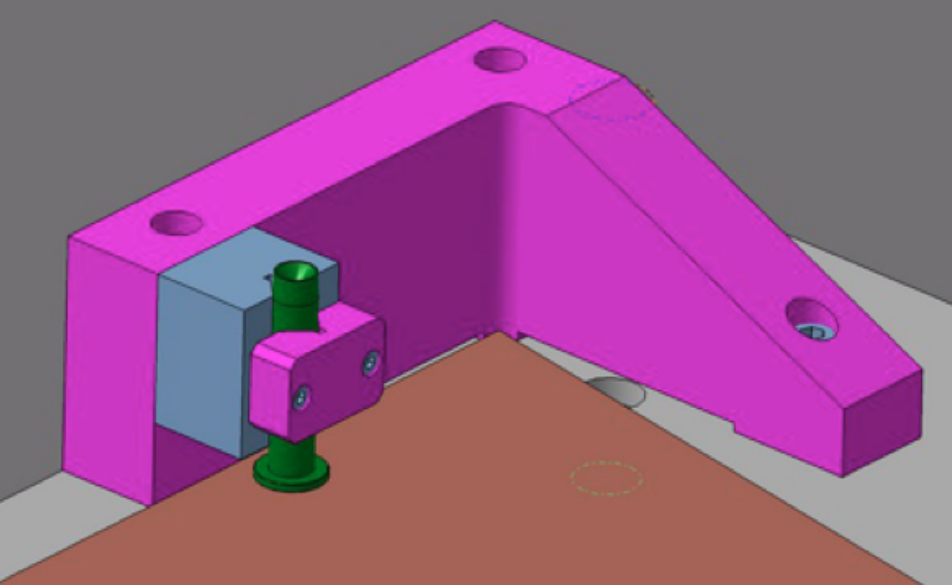}
	\caption{The tool to position the precision pin on
	 the lower readout panel with high acuracy and under
	 $90.0^{\circ}$. The tool will be removed after glueing.
	 The mounting position of the rigid clamps is given 
	 by precise pins in a stiffback or in a template.}
	\label{alignmentPinMounting_FIG}
\end{figure}

In addition to a precise diameter the pin also needs to be exactly 
perpendicular to the readout panel. 
In fig \ref{alignmentPinMounting_FIG} a tool for the correct pin 
adjustment is shown. 
The pin is glued to the panel vertically and on precise positions 
that are given by alignment pins and holes on stiffbacks or templates produced 
by precise CNC milling machines.
Before assembly all five panels must be leak tested using
an appropriate test facility.

\begin{figure}[h!]
	\centering
	\includegraphics[width=3.5in]{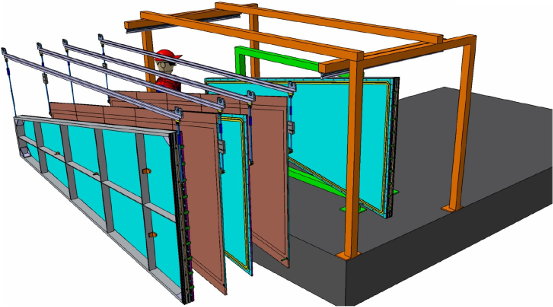}
	\caption{Assembly of a quadruplet in a vertical mounting structure. 
	The outer frames are reinforced by stiff frames which are removed 
	after the assembly. Vertical mounting in a clean room reduces 
	the influence of dust and dirt.}
	\label{assablyHanging_FIG}
\end{figure}

To minimize the contamination of the assembled quadruplet with dust and dirt, the final assembly takes place in a clean room and in a vertical
mounting structure
as shown in fig. \ref{assablyHanging_FIG}. 

The panel assembly process starts with one of the outer drift panels
mounted vertically to a solid holding frame.
The outer drift panels are mounted additionally on stiffening frames 
since they experience a torque
due to the 
stretched mesh which otherwise would 
cause bending of the panel.
 
The other panels are then guided by a linear bearing
system that allows for precise adjustment of the precision pins 
or the precision washers in a force-free manner monitored by load cells.
For the fine adjustment micrometer screws will be used.

Each time a drift and readout panel will be combined,
a HV test will be performed. This is a very sensitive test 
to remaining dust particles that have to be removed 
before final assembly.
After all panels are put together the quaduplet is fixed with screws. 
In a final step the external stiff frames are removed.

\section{Flatness of Sandwich Panels}
	\label{FoSP}

\begin{figure}[h]
	\centering
	\includegraphics[width=3.5in]{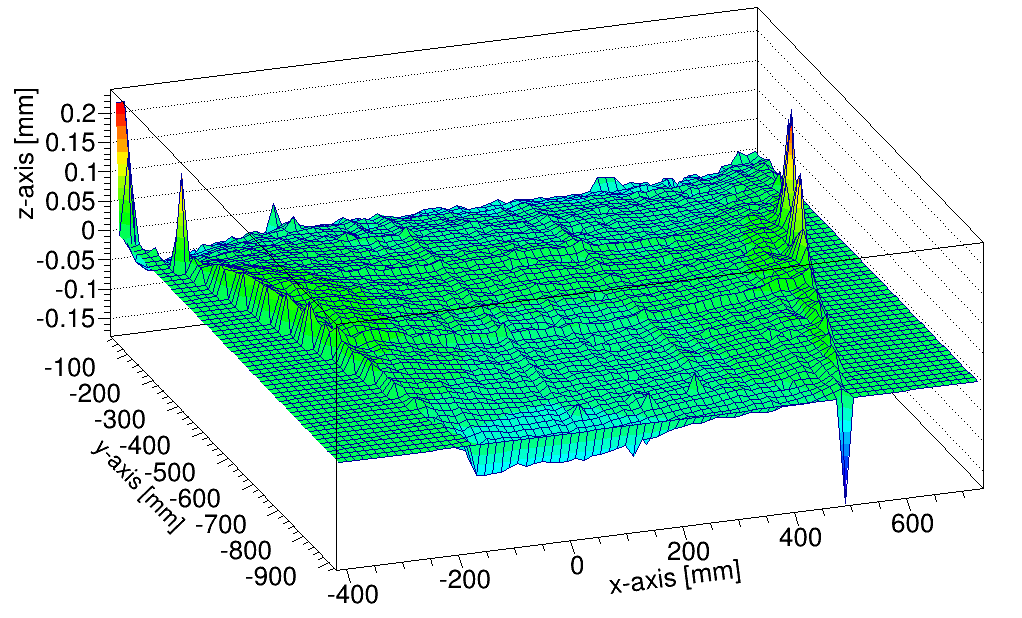}
	\caption{Topology of a mechanical prototype panel obtained 
	with a coordinate measurement machine. Deviations from the 
	tolerances occur only at the border, outside the active area.
	The 15\,$\mu m$ high grid like structure was intentionally provoked
	by grooves in the granite table. This allows easily for optical
	judgement of the planarity of the surface. The panels produced for the NSW will not show this grid like structure on their surfaces.}
	\label{topologyResult_FIG}
\end{figure}

To meet the stringent mechanical requirements for individual panels, 
for the whole quadruplet and finally for the whole sector,
the topology of the surfaces of a single panel is a figure of merit. 
Using either tactile or nontactile measurement methods,
or both, 
the flatness of each panel has to be mapped.
Fig. \ref{topologyResult_FIG} shows the topology of one side of a sandwich panel 
of the mechanical protoype which was built using the two step glueing method.

It shows an almost perfect flatness inside the active area of the %hypothetical
detector. Larger deviations are visible only
at the borders of the panel. 
Reasons for these deviations have been investigated and result 
in stiffbacks with a 10-fold higher stiffness than the one used 
for this prototype.

\begin{figure}[h]
	\centering
	\includegraphics[width=3.5in]{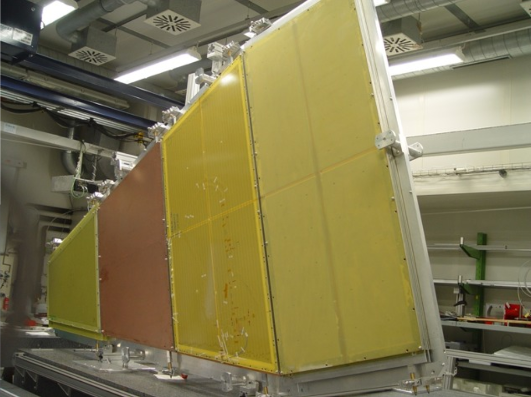}
	\caption{For deformation and stability tests 
	each of the four construction sites contributed a mechanical prototype
	built according to the structure of the final quadruplets. 
	Only the active readout anodes were substituted by dummies
	and the segmentation of a sector was chosen to be four-fold.
	This picture shows a full size model of a large sector built by four modules mounted on the spacer 
	frame in preparation for measurements as temperature dependence 
	of surface planarity or reversibility of deformations
	provoked by temperature, stress or overpressure.}
	\label{mechPrototype_FIG}
\end{figure}

Deviations from a planar surface are not only due 
to %a insufficient 
construction processes but can also result 
from operational conditions.
Fig.~\ref{mechPrototype_FIG} shows the assembly
of four mechanical prototypes on a spacer frame placed on a large granite table. 
Deformations by intentionally 
induced temperature gradients could be 
measured and simulated. 
They were largely understood and lead to a modified 
holding structure of the modules in the final setup.

\begin{figure}[h]
	\centering
	\includegraphics[width=2.8in]{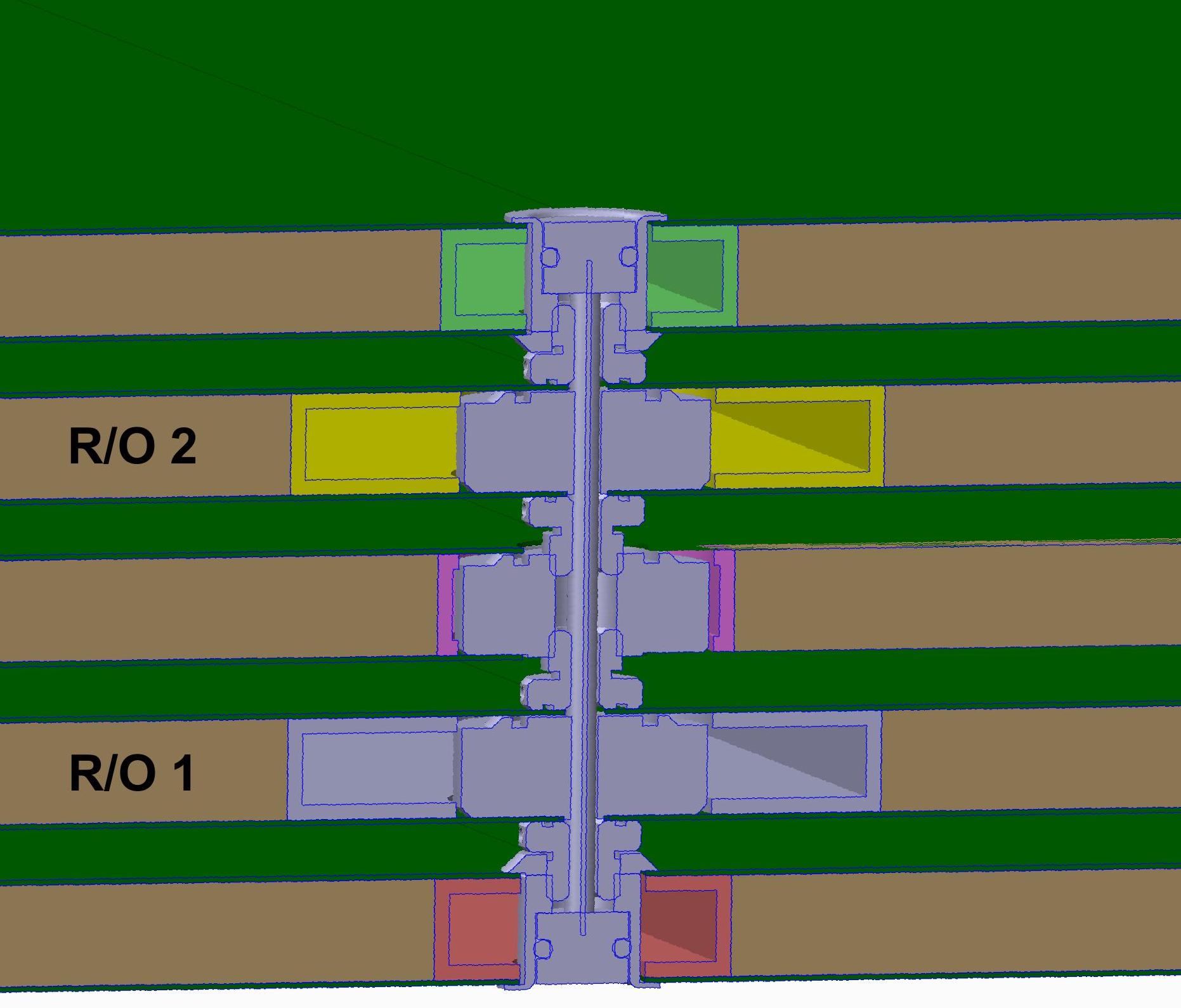}
	\caption{An interconnection through the five sandwich panels.
	The almost planar areas directly around the interconnections
	on the top side of the quadruplet will be used for the 
	precise positioning of alignment platforms for optical 
	monitoring of thermal deformations of the quadruplets during 
	active LHC runs.}
	\label{interconection_FIG}
\end{figure}

Since the detectors are filled with an overpressure 
of $2~mbar$ the whole quadruplet blows up during operation. 
To minimize the blow up interconnections are used 
which keep the thickness of the quadruplet constant. 
The layout of an interconnection is shown in fig. \ref{interconection_FIG}.

\begin{figure}[h]
	\centering
	\includegraphics[width=3.5in]{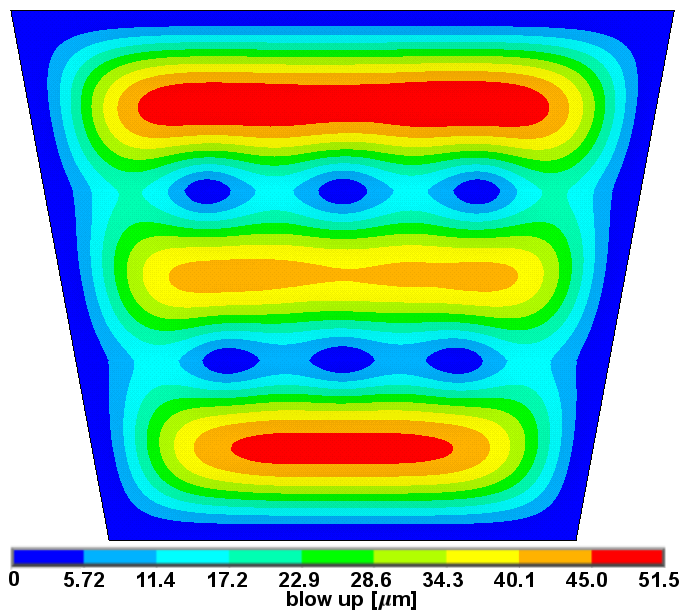}
	\caption{Simulation of blow up of quadruplet due to $2~mbar$ overpressure inside the chambers. This simulation shows the result using six interconnections. The maxmimum deviation is $\approx50~\mu m$.}
	\label{ansys_FIG}
\end{figure}

To find a minimal number and ideal positions of the interconnections 
simulations with the finite element code ANSYS \cite{ansys} are investigated. 
Fig. \ref{ansys_FIG} shows the result of a simulation 
with six interconnections in the SM2 module \cite{thesis_elias}.

The maximum deformations are with $\approx50~\mu m$ well below the 
maximum tolerance.

\section{Conclusion}
	\label{conclusion}
We describe the feasibility, the design and construction of 
high-rate capable large size Micromegas
chambers for the upgrade
of the ATLAS muon spectrometer.
To cope with the extreme requirements, 
as readout strip alignment within $30~\mu m$ 
and surface planarities within $80~\mu m$ precision,
special methods and tooling had to be developed
for these large detector modules having areas above 2\,$m^2$.
A basic idea is to transfer the planarity of 
precision surfaces, as granite tables, 
to the surface of the readout and drift panels.
These panels consist of sandwich structures
with surface sheets of three or five pieces of 0.5\,mm thick 
printed circuit boards 
and a core of aluminum honeycomb
for stability.
The drift panel PCBs form the cathodes 
and the readout panel PCBs the anodes of the 
detectors. 
The strip anodes are designed as resistive strips 
with specific resistivites around 10\,M$\Omega$ per cm.
The readout strips are coupled capacitively to 
the resistive strips and provide the signals for the 
front end electronics. Their  
strips are 300\,$\mu m$ wide
and have a pitch of 425 or 450\,$\mu m$  
for the small or large sectors, respectively. 
The width of the resisitive strips is slightly
smaller at identical pitch.

In this paper the working principle of resistive strip Micromegas 
detectors is explained. The design of the two New Small Wheels
is separated 
into eight sectors.
Each sector is subdivided into two trapezoidal subsectors.
Each subsector consists of eight consecutive detector layers 
separated by a spacer frame
into two quadruplets. 
The Micromegas quadruplets are built
from five sandwich panels separated by precise extruded aluminum
bars, the drift frames.
This is presented in detail including the vertical assembly procedure.
It is shown that the strict planarity requirements 
can be fulfilled.
Methods are presented that allow for precise and exact 
alignment of the readout PCBs against each other.
The stretching of meshes in the size needed for the NSW is possible. The cleaning of the meshes is still under investigation. 
Simulations have shown that the slight overpressure in the gas detectors leads to a deformation of the surface. 
To resolve this problem interconnection will be installed into the quadruplets, which keep the deformation to a minimum.
Then the maximal deformation due to the overpressure will be only $50~\mu m$.

The construction principle allows for a series production of many large area micromegas detector modules while ensuring a $\mathcal{O}\left(10~\mu m\right)$ level of mechanical accuracy for every module.

%\appendices
%\section{}
%Appendices, if needed, appear before the acknowledgment.

% use section* for acknowledgement
\section*{Acknowledgment}
We acknowledge the support by the DFG Excellence Cluster Universe.
%The preferred spelling of the word ``acknowledgment'' in American English is without an ``e'' after the ``g.'' 
%Use the singular heading even if you have 
%many acknowledgments. Avoid the expression, ``One of us (S.B.A.) thanks ...'' Instead, write ``S.B.A. thanks ...'' Put %sponsor acknowledgments in the unnumbered footnote on the first page.

% references section

% can use a bibliography generated by BibTeX as a .bbl file
% BibTeX documentation can be easily obtained at:
% http://www.ctan.org/tex-archive/biblio/bibtex/contrib/doc/
% The IEEEtran BibTeX style support page is at:
% http://www.michaelshell.org/tex/ieeetran/bibtex/
%\bibliographystyle{IEEEtran}
% argument is your BibTeX string definitions and bibliography database(s)
%\bibliography{IEEEabrv,../bib/paper}
%
% <OR> manually copy in the resultant .bbl file
% set second argument of \begin to the number of references
% (used to reserve space for the reference number labels box)

% that's all folks
\end{document}